\pdfoutput=1

\documentclass[a4paper,11pt]{article}

\usepackage{jinstpub} 

\usepackage{lineno}

\graphicspath{{./figures/}} 

\usepackage{mathtools}  

\usepackage{listings}

\usepackage{xcolor}

\newcommand{\Codify}[1]{{\footnotesize\texttt{#1}}}

\title{\boldmath The charge sensitivity calibration of the upgraded ALICE Inner Tracking System}

\author[a,1]{Shiming Yuan,\note{Corresponding author.}}
\author[a]{Johan Alme,}
\author[b]{Markus Keil,}
\author[b]{Ivan Ravasenga,}
\author[a]{Matthias Richter,}
\author[a]{and Dieter R{\"o}hrich}

\affiliation[a]{University of Bergen,\\Allégaten 55, 5007 Bergen, Norway}
\affiliation[b]{ALICE Collaboration, CERN \\Espl. des Particules 1, 1211 Meyrin, Switzerland}

\emailAdd{Shiming.Yuan@uib.no}

\abstract{
	The ALICE detector is undergoing an upgrade for Run 3 at the LHC. A new Inner Tracking System (ITS) is part of this upgrade. The upgraded ALICE ITS features the ALPIDE, a Monolithic Active Pixel Sensor. Due to IC fabrication variations and radiation damages, the threshold values for the ALPIDE chips in ITS need to be measured and adjusted periodically to ensure the quality of data. The calibration is implemented within the ALICE Online-Offline (O$^2$) Computing System, thus it runs in the same framework as the normal operations. This paper describes the concept and first implementation of the charge sensitivity scanning procedures for the upgraded ALICE ITS in the ALICE O$^2$ System, and demonstrates the first results of the scanning of the data taken from the installed ITS.
}


\collaboration[c]{on behalf of the ALICE Collaboration}

\proceeding{TWEPP 2021 Topical Workshop on Electronics for Particle Physics\\
  20-24 September, 2021\\
  Bergen, Norway}

\begin{document}
\maketitle
\flushbottom

\section{Introduction}
\label{sec:intro}

\begin{figure}[htbp]
	\centering
	\includegraphics[width=.8\textwidth]{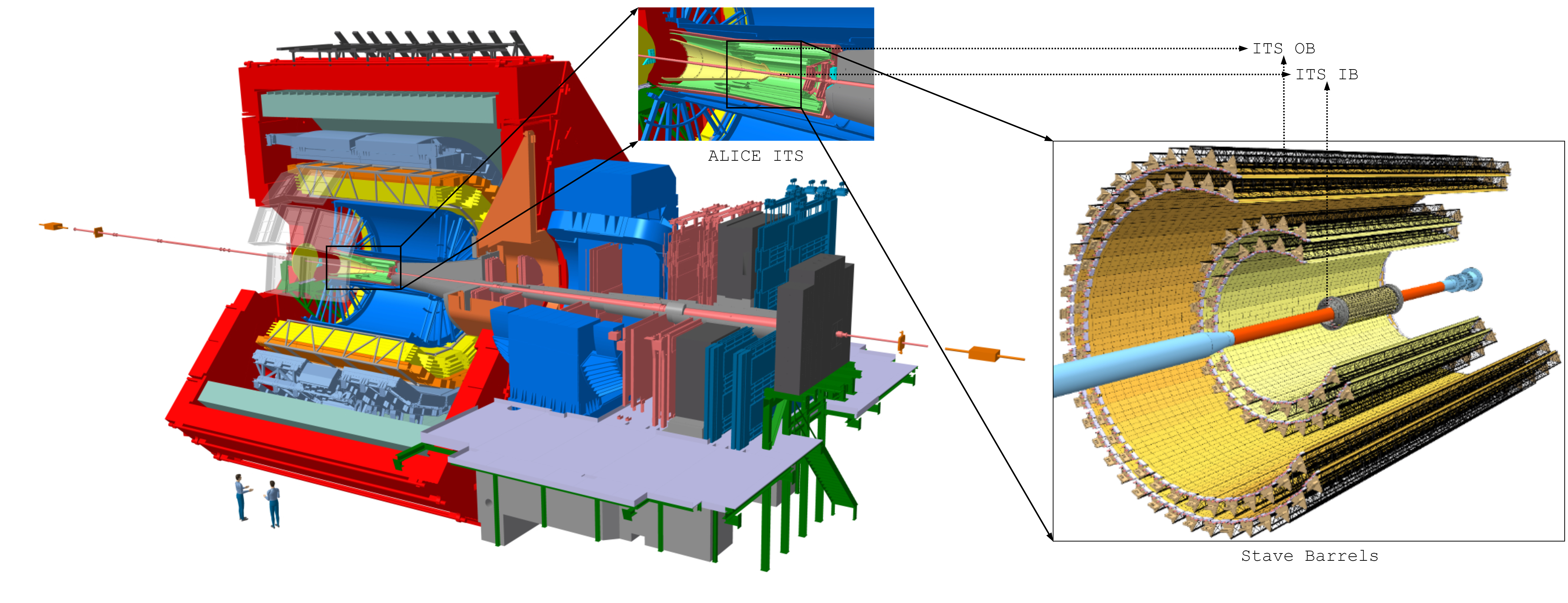}
	\caption{\label{fig-1} The ALICE Detector and the upgraded Inner Tracking System}
\end{figure}

The ALICE ITS (\textbf{I}nner \textbf{T}racking \textbf{S}ystem) is the innermost subdetector in the ALICE apparatus. The ITS is responsible of determining the primary vertices, reconstructing secondary vertices, and improving the resolution of the ALICE Time Projection Chamber\cite{contin2012performance}. A new ITS has been designed for the upcoming Run 3 at the LHC\cite{trzaska2020new}. 
The upgrade of the ITS features the ALPIDE (\textbf{AL}ice \textbf{PI}xel \textbf{DE}tector)), a custom Monolithic Active Pixel Sensor.
The upgraded ITS consists of seven concentric cylindrical layers of ALPIDE arrays surrounding the beam pipe at distances from the Interaction Point (IP) ranging from 22 mm at the inner most layer to 430 mm at the outer most layer. The location of the Inner Tracking System in ALICE is shown in Figure \ref{fig-1}.
The new ITS is currently in the final phase of testing and commissioning.

The ALPIDE chip contains a matrix of 512 (Y) $\times$ 1024 (X) sensitive pixels, and the periphery circuitry that handles both readout and control of the pixels on a 15 mm (Y) $\times$ 30 mm (X) dice, fabricated using the TowerJazz 0.18 $\mu$m technology\cite{mager2016alpide}. 
Due to variances in the fabrication processes and radiation damages\cite{lazareva2017analysis}, the threshold values of the ALPIDE chips need to be acquired, monitored, and tuned periodically to ensure the uniformity of data taking. 

This paper consists the following sections. First, the working princple of ALPIDE is outlined with an emphasis on the necessity and method of the threshold scanning. Second, the threshold scanning mechanism in the ALICE Online-Offline System (O$^2$) is described. Third, the results and current status of development are presented. The final section discusses the future work in the project. 

\section{The working principle of ALPIDE and S-curve measurement}
\label{sec:alpide}

\begin{figure}[htbp]
	\centering
	\includegraphics[width=.6\textwidth]{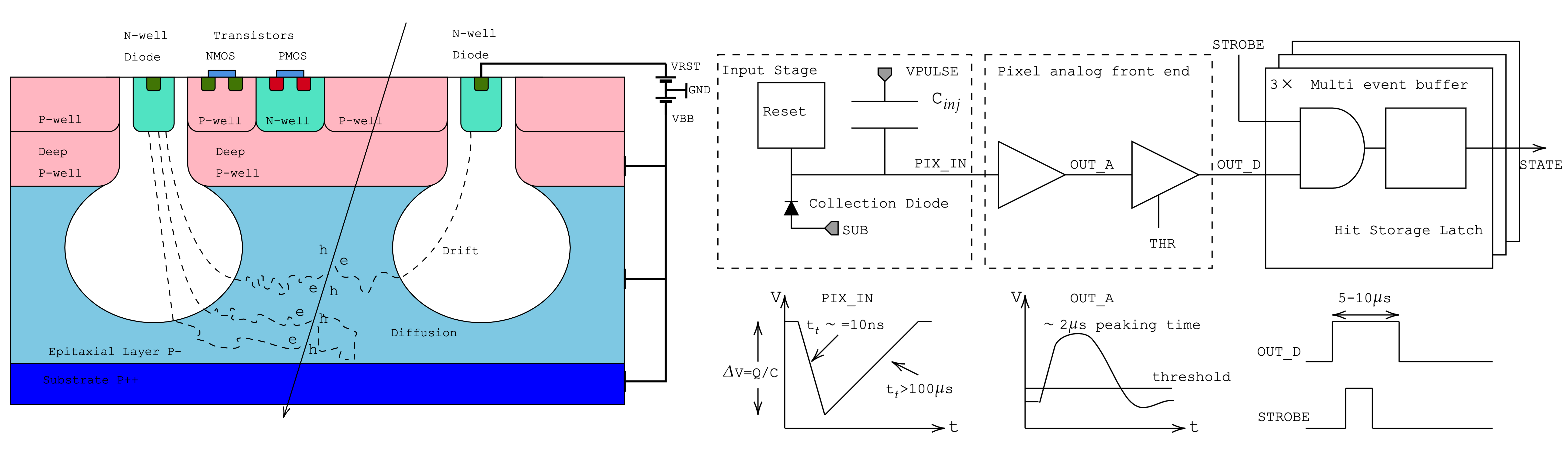}
	\includegraphics[width=.3\textwidth]{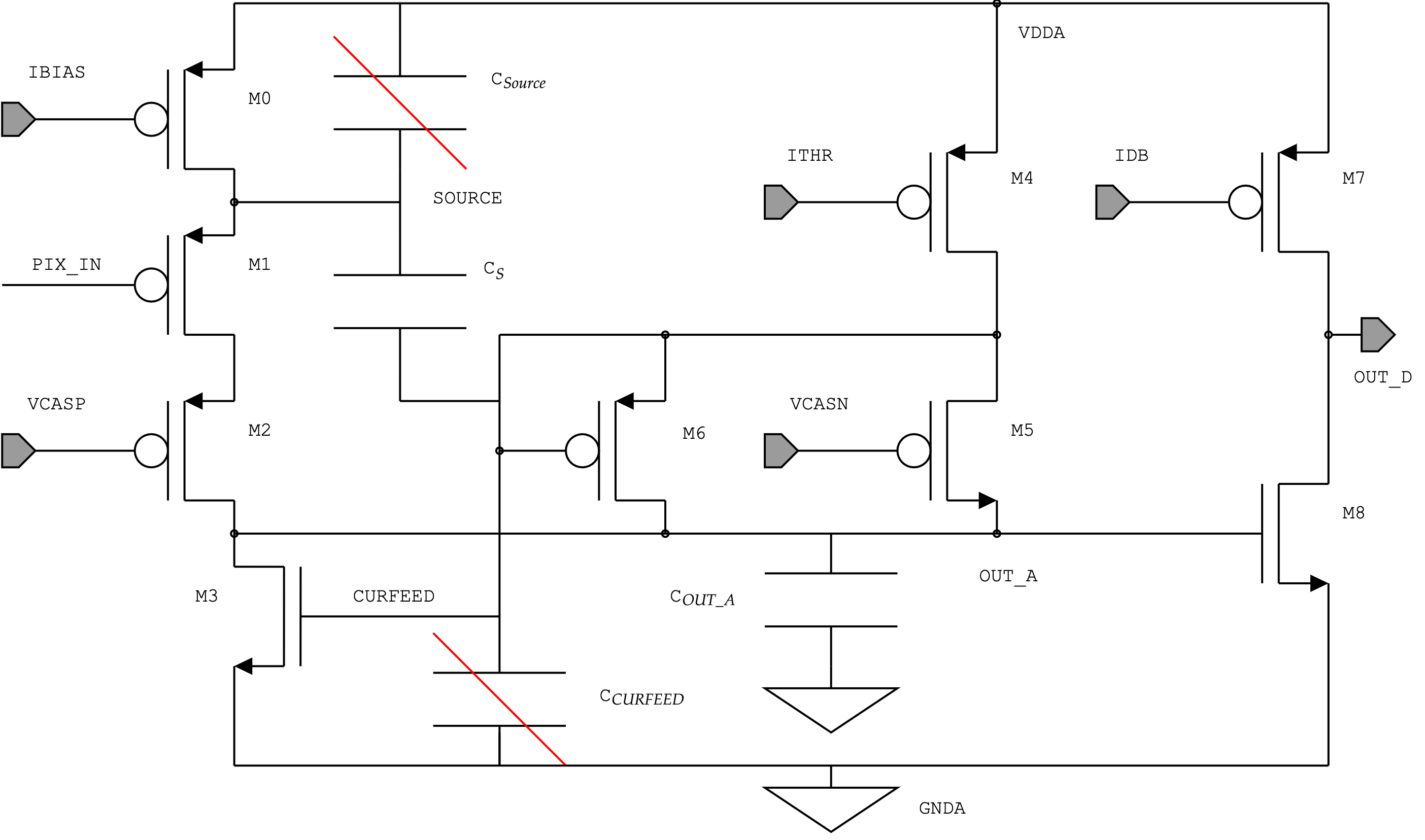}
	\caption{\label{fig-3} The cross section of ALPIDE, the signal processing to latch a hit, and the analog front end of a pixel}
\end{figure}

The cross section and signal processing diagram of ALPIDE are shown in Figure \ref{fig-3} \cite{mager2016alpide}. When there is a hit on the pixel, the electrons collected at the diode will induce a current signal at the input transistor (PIX$\textunderscore$IN). The analog circuit in the pixel amplifies this signal and if the amplified signal (OUT$\textunderscore$A) is greater than a configurable threshold, the signal OUT$\textunderscore$D will be passed to the digital multi event buffer, and latched as binary information.

One of the key parameters to monitor of the ALPIDE is the threshold that determines a particle hit on the pixel. The threshold value decides the amount of charge needed to generated a hit. The threshold value is set for all the pixels per chip.
However, due to radiation damage and fabrication variations, the threshold value for each pixel varies. 
Therefore it is critical to measure and calibrate the threshold values for all the ALPIDE chips in the ITS, in order to obtain uniform data by locating the pixels with wrong threshold and balancing the threshold setting for the whole chip.

While the charge threshold is defined by ITHR, VCASN, and IDB as shown in Figure \ref{fig-3}, the analogue values cannot be read directly. Thus the S-curve measurement is applied to read the threshold values. 

\begin{figure}[htbp]
	\centering
	\includegraphics[width=.25\textwidth]{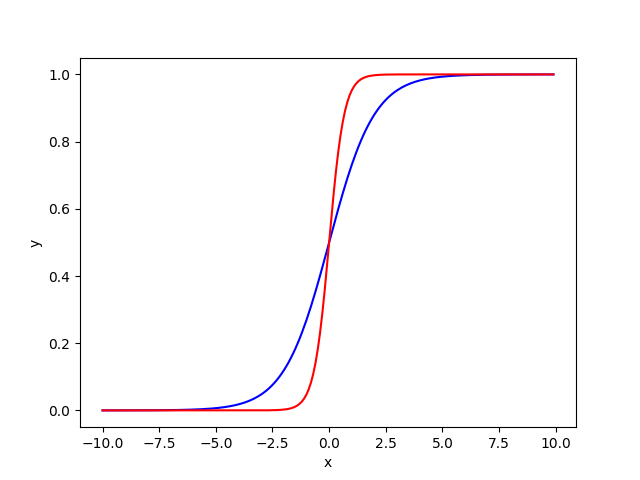}
	\includegraphics[width=.25\textwidth]{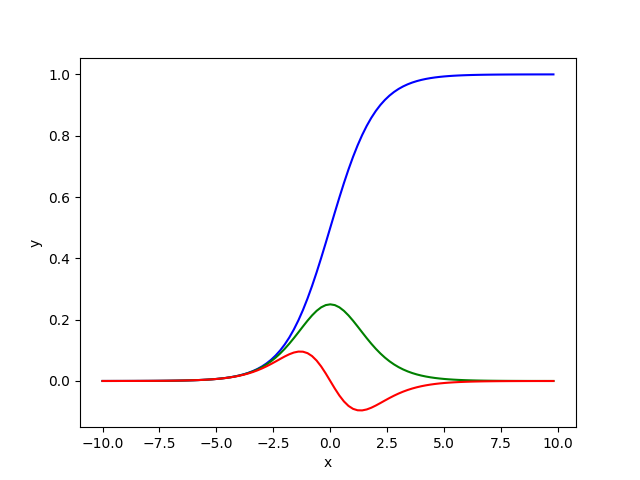}
	\caption{\label{fig:sigmoid} Shapes of logistics functions $y = \frac{1}{1+e^{(-x)}}$ (blue) and $y = \frac{1}{1+e^{(-3x)}}$ (red), and the first (green) and second (red) order derivatives of $y = \frac{1}{1+e^{(-x)}}$(blue)}
\end{figure}

The acronym ``S-curve'' originates from the sigmoid curve of different forms of continuous and differentiable approximations for the step activation function \cite{enwiki:1044853270}.
The shapes of two examples of logistic functions is illustrated in Figure \ref{fig:sigmoid}.

By utilizing the built-in analogue pulsing feature, discrete test charges can be injected to the pixel diodes. The threshold of the pixels can be measured by injecting a increasing series of discrete test charges. In theory, the only variable in an S-Curve scan is the C$_{inj}$ show in Figure \ref{fig-3}. The calibration runs will provide data sets containing the discrete values of C$_{inj}$ and the fired pixel ratio, thus the threshold value of a certain chip could be found by running a S-curve fitting for the discrete data. 
 
 Finding the threshold and sensitivity out of the data points is equivalent to finding the Sigmoid function $y = \frac{1}{1+e^{(-A(x+B))}}$ that fits the data set, where A reflects the steepness of the curve, or the sensitiveness of the pixel, and B represents the threshold value. Since the first order derivative of the $y = \frac{1}{1+e^{(-A(x-B))}}$ is $y' = \frac{Ae^{-A(x-B)}}{(1+e^{-A(x-B)})^2}$, $y'(B)=\frac{A}{4}$, so $A = 4 \times y'(B)$. And the threshold B can be found by finding the maximum of the first order derivative of the function.
 
\begin{figure}[htbp]
	\centering
	\includegraphics[width=.3\textwidth]{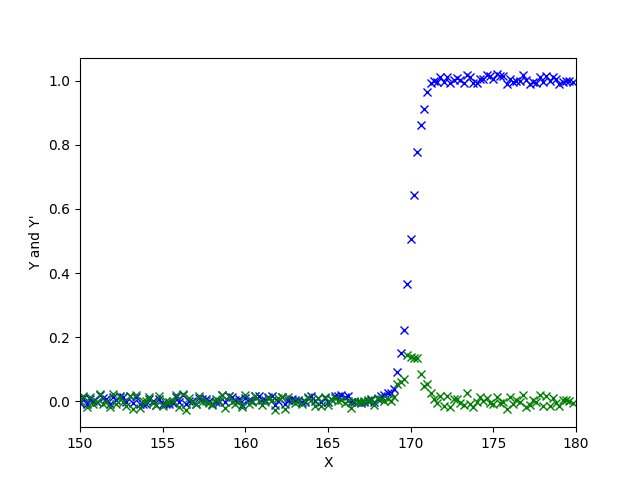}
	\caption{\label{fig:sigmoid-sim} Simulated data points for S-curve fitting}
\end{figure}

 For instance, the blue crosses in Figure \ref{fig:sigmoid-sim} illustrate the data points simulated by adding a random noise to $y = \frac{1}{1+e^{(-3(x-170))}}$. So the data set is $\{Y_N = \frac{1}{1+e^{(-3(X_N-170))}}+Noise| X_N=\{150, 151, ... , 180\}\}$. The green crosses represent array $\{Y'_N | Y_N - 0, \text{if} \, N = 150; Y_N - Y_{N-1}, \text{if} \, N \in \{151, 152, ... , 180\}  \}$.

By finding the maximum value in $\{Y'_N\}$, $A$ and $B$ can be determined, thus forms the S-curve.

\section{Threshold scanning in the ALICE O$^2$}
\label{sec:threshold_scanning}

\begin{figure}[htbp]
	\centering 
	\includegraphics[width=.5\textwidth]{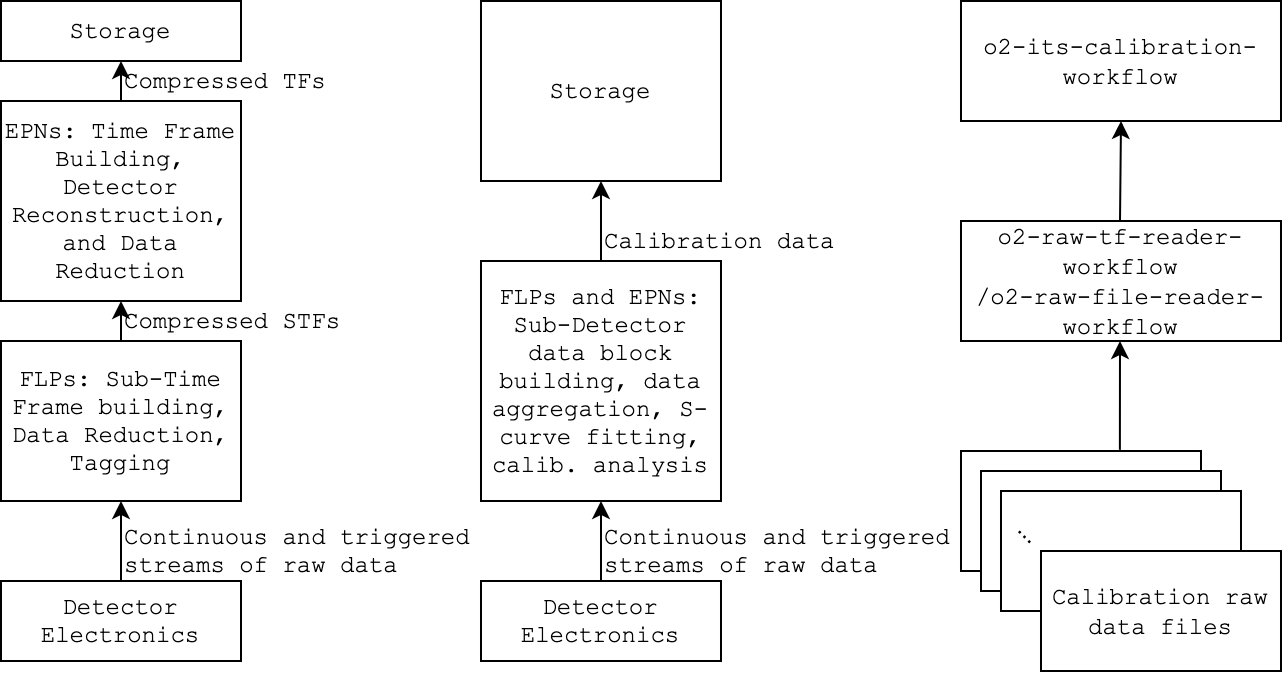}
	\caption{\label{fig:procedure} Normal analysis hierarchy, calibration hierarchy \cite{buncic2015technical}, and calibration workflow}
\end{figure}

The calibration algorithms differ from the normal data analysis primarily in the fact that in normal analysis, data of the same time stamp (Time Frame) are collected, whereas due to the inherent spatial nature of calibration, the calibration algorithms accumulate and analyze the data based on the spatial attributes. The minimal calibration workflow is illustrated in Figure \ref{fig:procedure}.

The scanning of the ALPIDE threshold values is done via a pipeline of workflows in the ALICE O$^2$ framework. The workflows define algorithms in the DPL (Data Processing Layers), and carry out different steps of handling of the raw data. For the calibration of the ITS, several existing workflows have been utilized. The \Codify{o2-raw-file-reader-workflow} and the \Codify{o2-raw-tf-reader-workflow} both parse the raw data in different formats and passes the parsed data as DPL messages to the next workflow in pipeline. The \Codify{o2-raw-file-check} validates the correctness of the data format, in particular the Raw Data Header version. For the threshold scanning, the \Codify{o2-its-calibration-workflow} is created.
In this workflow, multidimensional arrays like {\Codify{int HitMap[Charge\_inj][Row][Col]}} are created to store the hit-map parsed from the calibration raw data. And S-curve measurement algorithms described in Section 2 are implemented in this workflow to calculate the threshold value and the sharpness of the chips. 

\begin{figure}[htbp]
	\centering 
	\includegraphics[width=.25\textwidth]{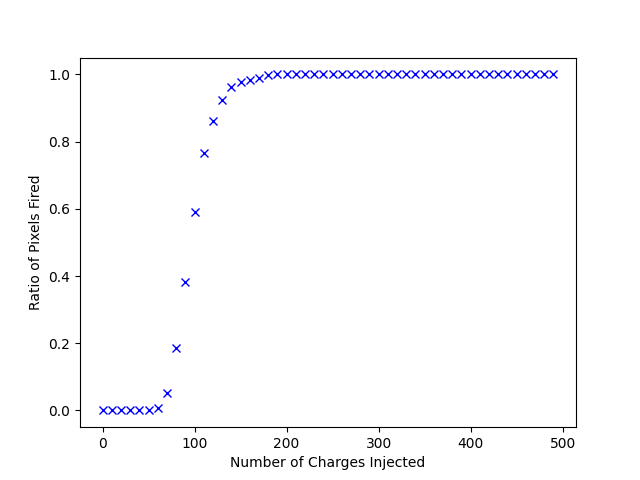}
	\includegraphics[width=.25\textwidth]{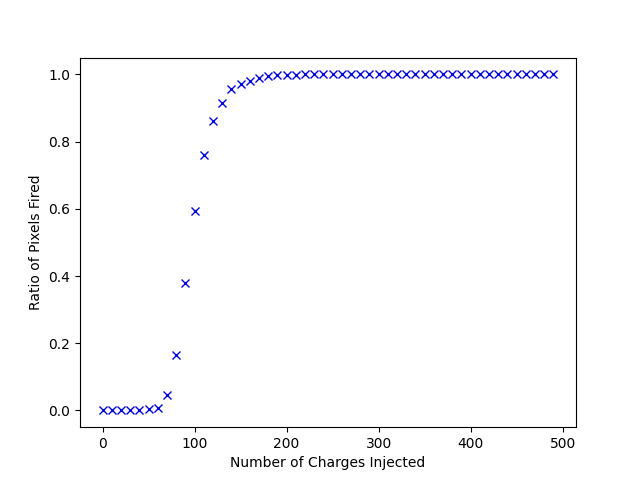}
	\includegraphics[width=.25\textwidth]{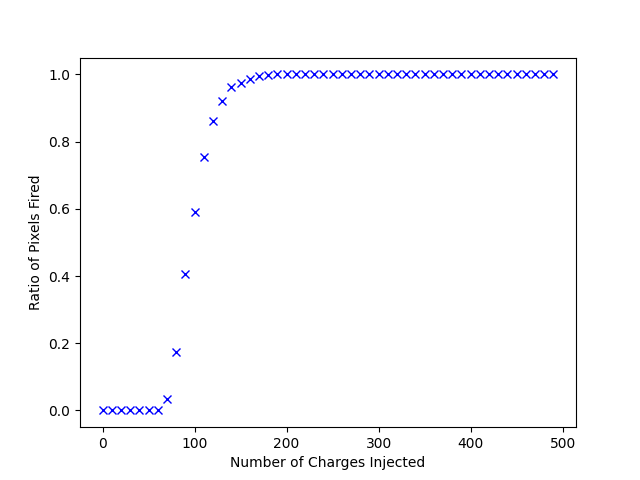}
	\includegraphics[width=.25\textwidth]{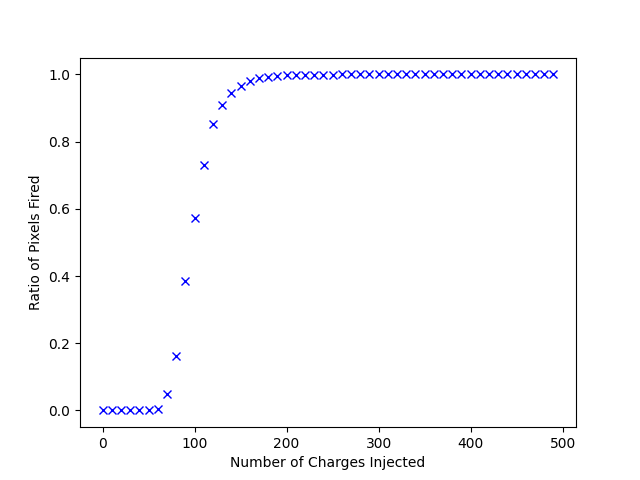}
	\includegraphics[width=.25\textwidth]{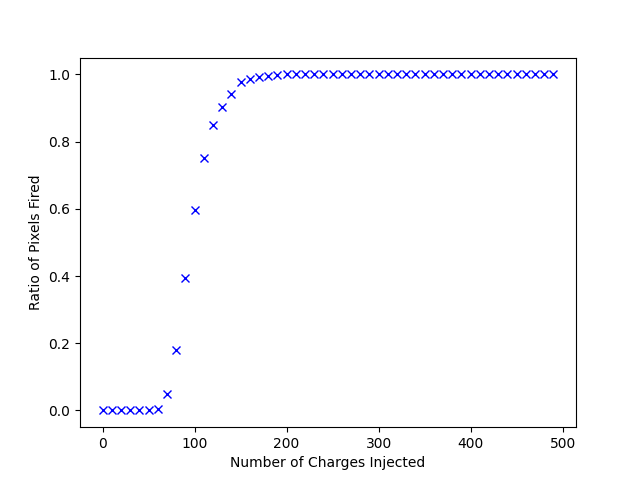}
	\includegraphics[width=.25\textwidth]{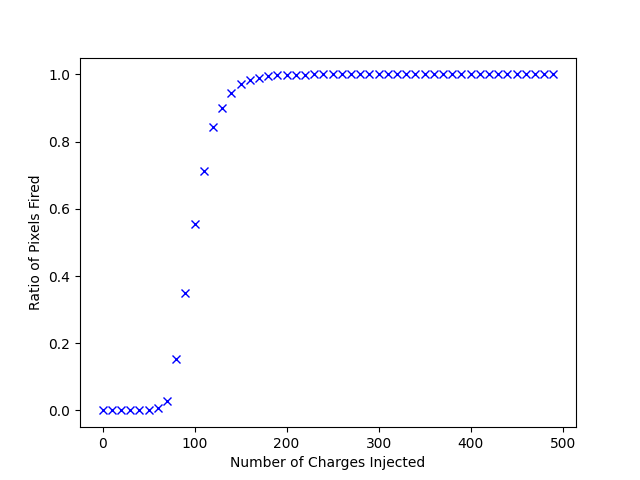}
	\caption{\label{fig:test_data_chip63_rows} Fire ratio of pixels on Rows 10 to 15, Chip 63 from sample calibration data set}
\end{figure}

\section{Results}

\begin{figure}[htbp]
	\centering 
	\includegraphics[width=.25\textwidth]{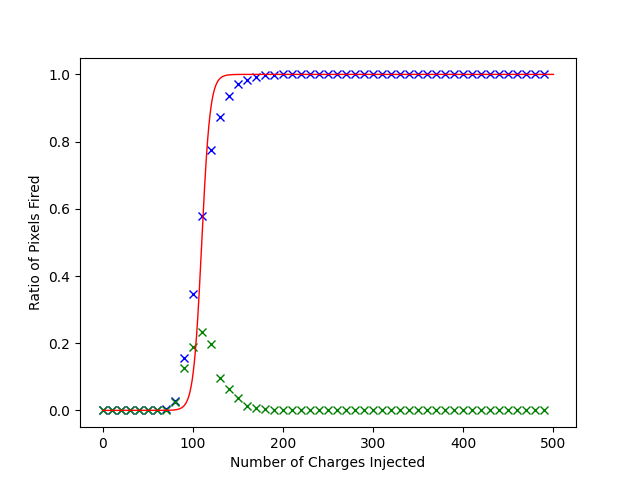}
	\includegraphics[width=.25\textwidth]{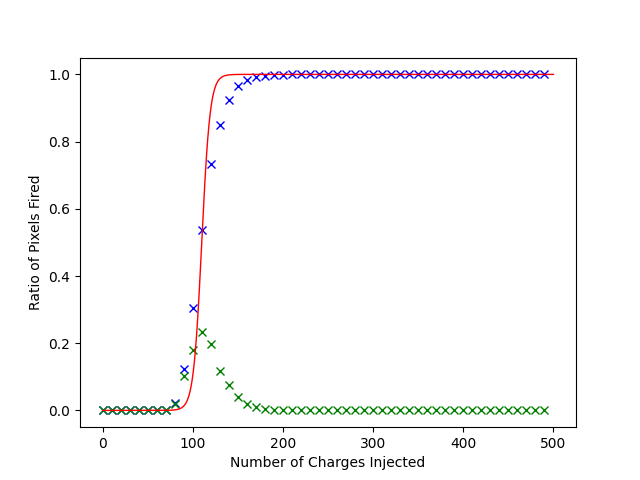}
	\includegraphics[width=.25\textwidth]{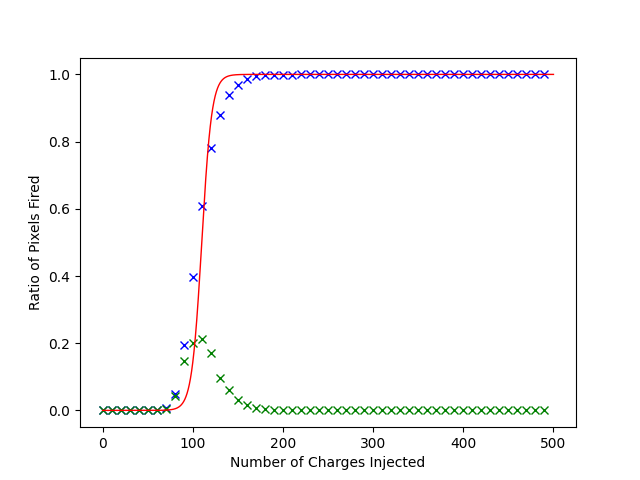}

	\caption{\label{fig:fit} S-Curve fitting for sample calibration data}
\end{figure}

Dedicated test runs have been performed at CERN to produce sample data for testing the calibration components. The development and results described in the paper are based on the v21.15 release of the ALICE O$^2$ repository, and the version 6 of the Raw Data Header format. The charge sensitivity calibration prototype described in this paper are also evolving with the update of the ITS firmwares and the O$^2$ computing system. The ITS calibration workflow extracts the run settings (number of charges injected) and the hit information (which pixels are fired under certain run setting) from the raw data. The \Codify{HitMap} data structure allows examination of the calibration data from different granularities. Figure \ref{fig:test_data_chip63_rows} illustrates the hit ratio of Rows 10 to 15 of Chip 63 from a calibration data set taken on 28 September, 2021. Figure \ref{fig:fit} illustrates the hit ratio of Chips 420, 421, and 422 from the sample calibration data set mentioned above, only rows 10 to 20 were fired for this data set. As described in Section 2, Figure \ref{fig:fit} further illustrates the process of finding the logistic function that fits the plot. A threshold value of 110 electrons was found for the 3 chips in the sample data set.

\section{Outlook}

The ALICE ITS apparatus and the ALICE O$^2$ are both undergoing testing and updating after the installation in May, 2021. The calibration task force will take more data set to test the software described in this paper. The software will need to be incorporated with proper ways of visualization and storage.



\end{document}